\documentclass{webofc}

\usepackage[varg]{txfonts}   
\usepackage{hyperref}
\usepackage{url}
\hypersetup{colorlinks=true,citecolor=blue,urlcolor=blue,linkcolor=blue}
\begin{document}
\title{Quantum computing for heavy-ion physics: near-term status and future prospects}
%
%

\author{\firstname{Jo\~{a}o} \lastname{Barata}\inst{1}\fnsep\thanks{\email{joao.lourenco.henriques@cern.ch}} 
}

\institute{CERN, Theoretical Physics Department, CH-1211, Geneva 23, Switzerland
          }

\abstract{We discuss recent advances in applying Quantum Information Science to problems in high-energy nuclear physics. After outlining key developments, open challenges, and emerging connections between these disciplines, we highlight recent results on the study of matter states, hard probes, and spin correlations using novel quantum technologies. This work summarizes the corresponding presentation delivered at the Quark Matter 2025 conference in Frankfurt, Germany.}
\maketitle
\section{Introduction}\label{sec:intro}
\begin{figure}[h!]
    \centering
    \includegraphics[width=1\linewidth]{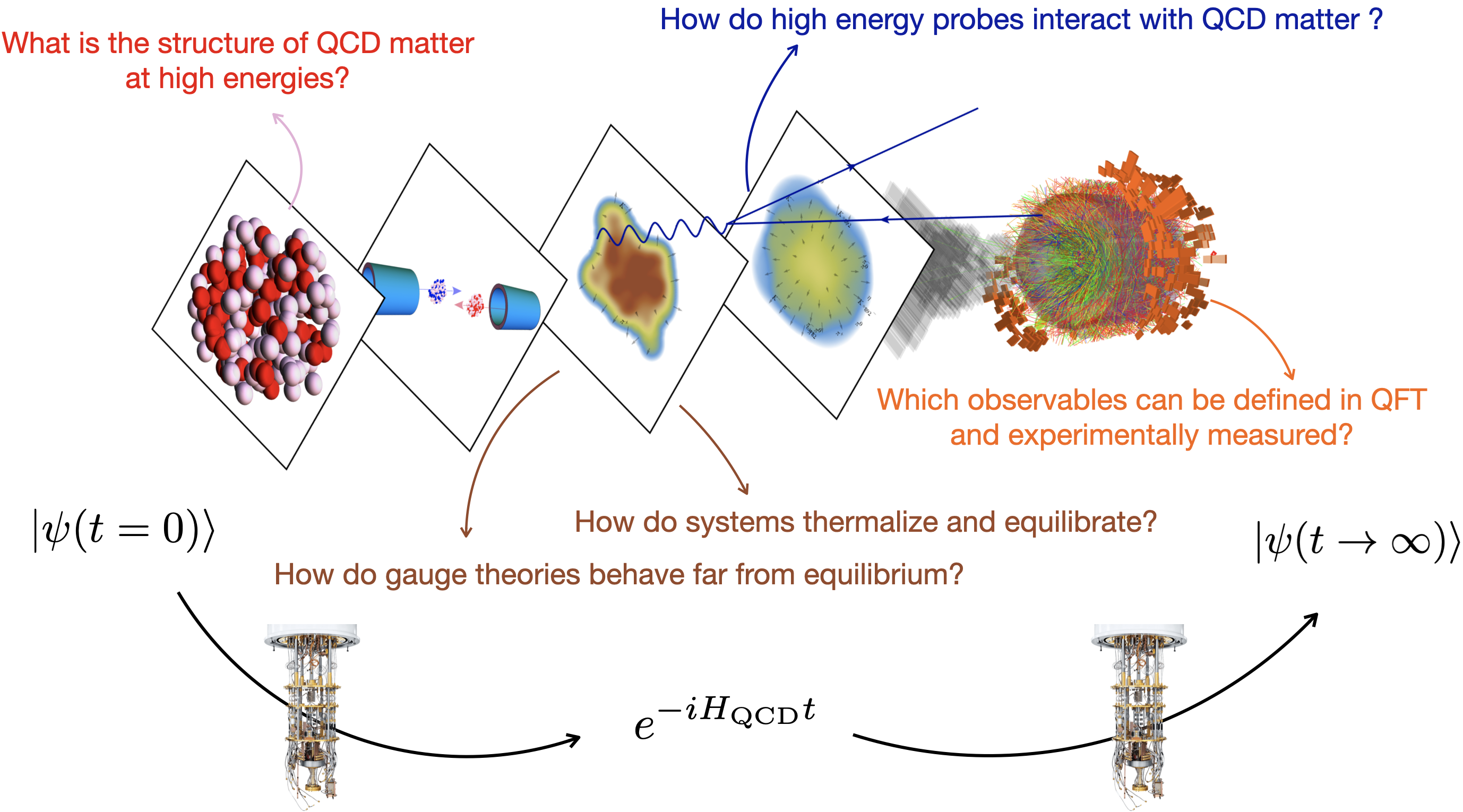}
    \caption{Diagrammatic representation connecting the standard theoretical picture for the evolution of the bulk matter in heavy ion collisions (top) with the real-time evolution possible to implement in a quantum computer (bottom).}
    \label{fig:cartoon}
\end{figure}

Ultrarelativistic high energy collisions of heavy ions have provided a unique experimental platform to explore the finite temperature and density, and out-of-equilibrium properties of QCD. Perhaps most remarkably, at the beginning of the century, this experimental effort resulted in the production of quark gluon plasma (QGP) droplets at RHIC, a state of matter similar to the one that populated the early universe where quarks and gluons are deconfined from hadrons, see~\cite{Busza:2018rrf} for a recent review. Since then there has been a large effort in mapping out the properties of this matter state, with perhaps the most important discovery being its small viscosity to entropy ratio, making it the most perfect fluid in the Universe. Despite this and other towering achievements, several questions about the evolution of QCD matter during heavy ion collisions remain open both on the experimental and theoretical fronts --- some of these are briefly summarized in Fig.~\ref{fig:cartoon}.

In part, the challenge in addressing some of the questions of interest, such as the transport and dynamical properties of the bulk or structure of the phase diagram of QCD, lie with their inherent real-time, finite density or non-perturbative character. As a result of the former two, lattice methods, which allow to study (euclidian) QCD from first-principles, are not applicable or are severely affected by noise problems; for the latter, perturbative approaches are incapable of providing any answer. Although there are several methods to overcome these challenges, in this summary article we discuss how advances in Quantum Information Science (QIS) and related fields can be used to surpass these obstacles and leverage our understanding of dynamical properties of QCD and quantum field theory in general. 

In what follows, we give first a brief overview of the current status in the description of heavy ion collisions. We follow to summarize recent progress in QIS methods and tools, and justify why they should be of interest to the heavy ion community. We elucidate this connection by providing several examples of applications which interface these two communities. We conclude this article by commenting on some future directions of research and possibilities that are left to be explored, before summarizing the main points of the discussion.

\section{The landscapes}\label{sec:landscape}

\subsection{The heavy ion physics landscape}\label{sec:HI_landscape}
A simplified representation of the current (theoretical) understanding of the time evolution of the bulk matter in the aftermath of a heavy ion collision is shown in Fig.~\ref{fig:cartoon}. In simple terms this can be decomposed into four different epochs: \textbf{1)} an initial state of two incoming nuclei, which are typically described in the context of the Color Glass Condensate~\cite{Gelis:2010nm}; \textbf{2)} the first instances ($\tau\leq 1$ fm/c) after the collision are characterized by the formation of a far-from-equilibrium QCD state which expands and equilibrates~\cite{Baier:2000sb} before reaching; \textbf{3)} a hydrodynamically evolving quark gluon plasma, where quarks and gluons escape the hadronic confinement; \textbf{4)} at late times ($\tau \geq 5 $ fm/c), after the bulk matter has reached the critical temperature and the confinement mechanism becomes dominant, the bulk matter converts to a free streaming collection of 
hadrons which are experimentally detected. A detailed summary of all these stages is given in e.g.~\cite{Berges:2020fwq, Heinz:2013th}.

Without diving deeper into the details of these stages, the common aspect to all of them is the need for a dynamical understanding for the formation of matter and the relevance of non-perturbative physics. As a result, each one of these stages is typically described within an effective description (kinetic theory, relativistic hydrodynamics, hadronic gas, $\cdots$). Although such approaches have been phenomenologicaly successful several conceptual questions remain open, such as how can one interconnect descriptions which have different basic degrees of freedom, or how can such effective theories be connected to QCD. These questions are not only interesting from the theory point of view, but making progress towards answering them might give clues towards experimental observables which could test whether the current picture for the evolution of the bulk matter is indeed realized in Nature. So far this last point has not been fully tested in experiment, opening the possibility that alternative descriptions of the matter evolution could hold, see e.g.~\cite{Bierlich:2016vgw}.

\subsection{The QIS landscape}\label{sec:QIS_landscape}
QIS has seen an accelerated development over the past decades, with the emergence of new technologies, such as quantum computers and simulators, and new methods, such as more efficient quantum algorithms, and related classical methods as tensor networks, see e.g.~\cite{Banuls:2019bmf, Banuls:2018jag, Jordan:2012xnu,Gilyen:2018khw}. 

Computational QIS methods offer a paradigmatic new approach to solving many-body dynamics in quantum systems: they allow for the simulation of the exact dynamics of a reference physical system, by either evolving another quantum state (quantum computing) or by efficiently representing it in a classical machine (tensor networks), without incurring in classic sign problems. These allows to tackle real-time and finite density processes, which are out of reach for standard euclidean lattice methods. Nonetheless, currently the available quantum platforms for simulations of gauge theories are still limited, see~\cite{Cobos:2025krn,Gonzalez-Cuadra:2023rex} for some recent developments, and most focus has been put on two dimensional theories, where classical tensor network methods can already perform accurate and large simulations, see e.g.~\cite{Banuls:2013jaa, Magnifico:2024eiy,Magnifico:2020bqt}.

To better illustrate how QIS could help in the description of heavy ion collisions, on the bottom of Fig.~\ref{fig:cartoon} we depict how real-time time evolution of the bulk matter could be, in principle, emulated by a quantum computer. For that, one would first prepare a quantum state where two boosted wavepackets approach. The time evolution of this system is naturally realized in the quantum computer, provided the QCD Hamiltonian can be implemented. At asymptotic late times, the collection of hadrons seen in experiment should be captured by the quantum state prepared in the quantum computer; this can be sampled by making measurements on the quantum device. The key point is that the entire simulation only involves unitary evolution via the QCD Hamiltonian; thus this is a fully non-perturbative, first-principle and real-time approach, which can help benchmark the theoretical effective descriptions of the bulk matter.

\begin{figure}[h!]
    \centering
    \includegraphics[width=0.8\linewidth]{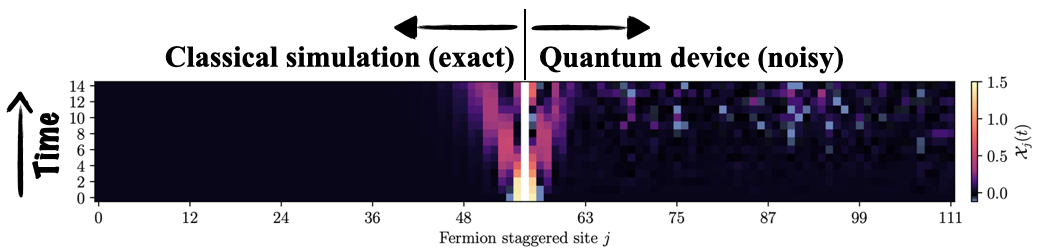}
    \caption{Wavepacket evolution, visualized through the local chiral condensate $\chi_j$, using exact (classical) simulation (left) and quantum simulation in a real device (right). Figure adapted from~\cite{Farrell:2024fit}.}
    \label{fig:wvps}
\end{figure}

Nonetheless, the actual realization of such a program is still most likely a long term goal, and rather the current focus should be put on solving simpler/smaller scale questions. To better illustrate the limitations of current quantum platforms, in Fig.~\ref{fig:wvps} we show the time evolution of a meson wavepacket prepared in two dimensional QED using both exact (classical) methods [left] and the equivalent simulation performed in state of the art quantum processor from IBM [right]~\cite{Farrell:2024fit}. The quantum result, even for this rather simple process, is quickly dominated by noise associated to the imperfect nature of the current quantum devices, while the classical method can easily perform such a simulation.

\section{Near-term QIS applications in nuclear physics}\label{sec:applications}
In this section we present some recent applications of QIS in the high energy nuclear physics context. 

\begin{figure}[h!]
    \centering
    \includegraphics[width=0.6\linewidth]{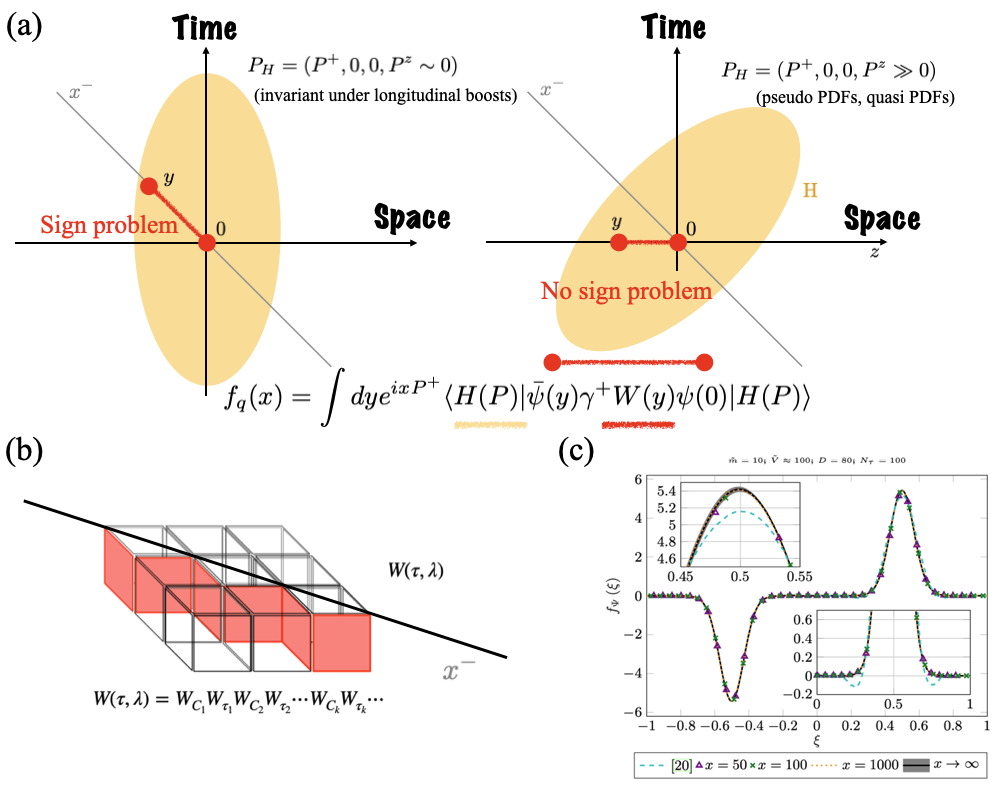}
    \caption{\textbf{(a)} Illustration of the computation of the matrix element entering the PDF for a quark $f_q$: on the left we have the formulation involving out-of-time correlations of the quark fields connected by a light-like Wilson line, the right figure illustrates the formulation from~\cite{Ji:2013dva,Radyushkin:2017cyf}, which allows to extract the PDF from space-like correlators. \textbf{(b)} Discretization of light-like Wilson lines in a form ammeanable for quantum computations~\cite{Echevarria:2020wct}. \textbf{(c)} Application of the strategy in panel \textbf{(b)} to the calculation of the PDF for the first excited state in two-dimensional QED~\cite{Banuls:2025wiq}.}
    \label{fig:pdf}
\end{figure}

\noindent \paragraph{Nuclear structure:} One of the most promising applications for QIS techniques lies in the characterization of quantum states of matter. In the QCD context, these could allow for the direct calculation of matrix elements involving out-of-time correlations of fields, not directly accessible in euclidean lattice QCD computations. This is illustrated in Fig.~\ref{fig:pdf} (a) for the case of the distribution $f_q(x)$ computed in a hadron state with momentum $P$, and involving a non-local quark billinear represented on the left panel in red, with $W(y)$ the light-like Wilson line needed to ensure gauge invariance. Here $x$ can be understood as the energy fraction carried by a \textit{quark} in the hadron.  In QCD this object can be identified with the bare parton (quark) distribution function (PDF) typically entering collinear factorization theorems~\cite{Collins:1981uw}. In euclidean lattice QCD such objects can be extracted after mapping the real-time correlator to an equal-time one, see~\cite{Ji:2013dva,Radyushkin:2017cyf} (see right panel); however this requires a matching to an appropriate ultraviolet description which must be computed separately, see~\cite{Boussarie:2023izj} for more details.

The quantum computation of out-of-time and non-local gauge invariant correlation functions is still, in practice, challenging to perform using QIS or related methods. Nonetheless, there have been developments in the calculation of extended gauge invariant objects, such as Wilson loops and Wilson lines connecting quarks field~\cite{Echevarria:2020wct,Zohar:2021wqy,Zohar:2015hwa}; in Fig.~\ref{fig:pdf} (b) [adapted from~\cite{Echevarria:2020wct}] we show a realization suited for quantum devices of a Wilson line operator, involving a \textit{ladder} discretization along the light-like direction. This discretization interpolates between real time evolution and the injection of electric string operators for the spatial component. This strategy has been recently applied in the calculation of the PDF in (1+1)D QED on the first excited state of the theory~\cite{Banuls:2025wiq}; the results of this work are summarized in Fig.~\ref{fig:pdf} (c), where the expected peak at energy fraction $x=0.5$ is found for the Schwinger boson state. Related work on PDFs and hadronic structure can be found in e.g.~\cite{Barata:2024bzk,Mueller:2019qqj,Lamm:2019uyc,Kreshchuk:2020dla,Li:2021kcs,Grieninger:2024cdl,Kang:2025xpz,Qian:2021jxp, Bauer:2021gup,Chen:2025zeh}.

\noindent \paragraph{Quarkonia, hard probes and energy loss:} 
Another avenue where QIS can provide new insights relates to the real-time dynamics of highly energetic or massive states propagating in the presence of bulk matter, as is observed for jets, quarkonia or open heavy flavor produced in heavy ion collisions as illustrated in Fig.~\ref{fig:cartoon}. Although there has been ample progress in the theoretical description of such probes in QCD~\cite{Apolinario:2022vzg}, there are still many simplifying assumptions, such as neglecting back-reaction on the bulk from the probe or the reliance on perturbation theory, which can be lifted using quantum simulation methods.

\begin{figure}[h!]
    \centering
    \includegraphics[width=0.6\linewidth]{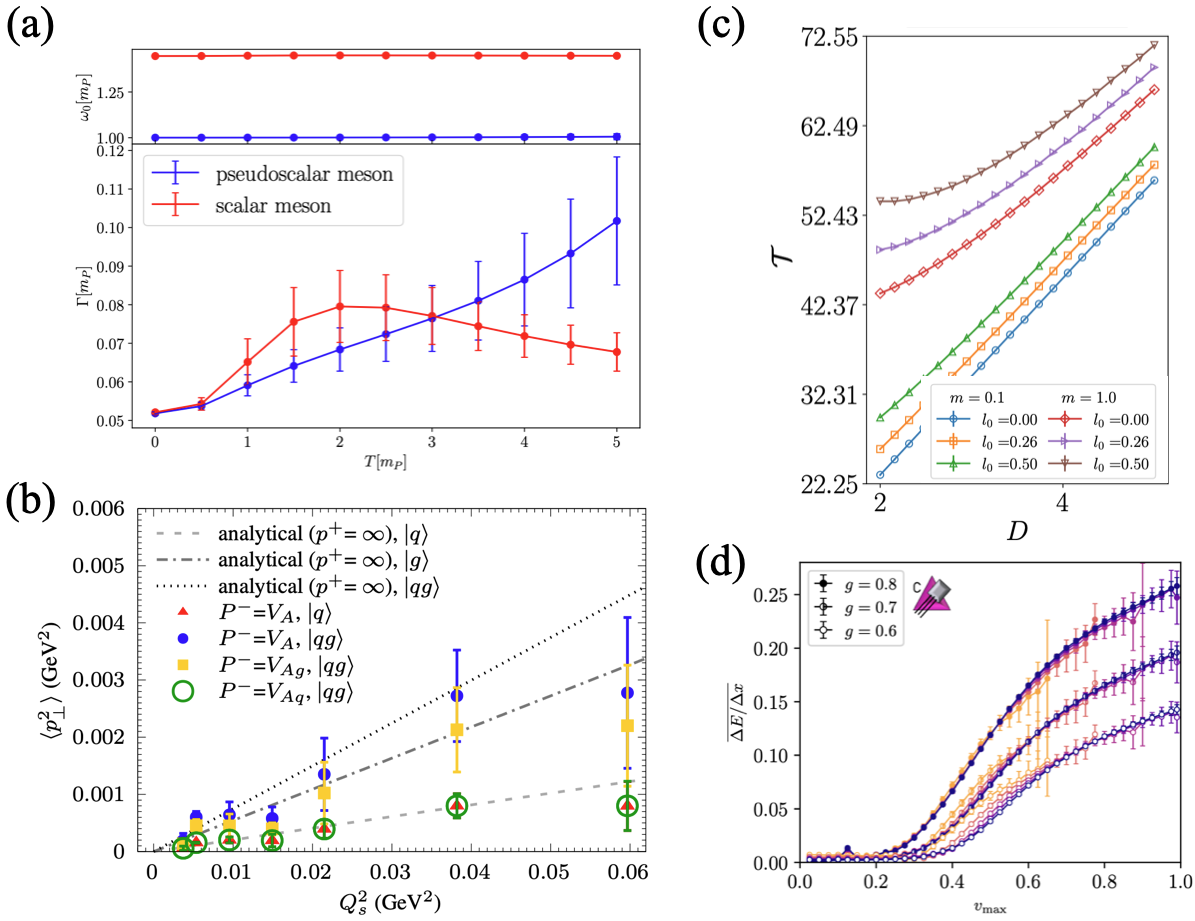}
    \caption{Examples of quantum simulation of the dynamics of heavy and energetic probes immersed in a matter background: (a) evolution of the spectral functions of mesons at finite temperature in two dimensional QED~\cite{Barata:2025jhd}, (b) average momentum broadening of jets as a function of the saturation scales of the medium~\cite{Barata:2023clv}, (c) thermalization time for mesons as a function of drag in a two dimensional QED~\cite{Angelides:2025hjt}, (d) energy loss for a heavy particle in two dimensional QED~\cite{Farrell:2024mgu}.}
    \label{fig:jets}
\end{figure}

In Fig.~\ref{fig:jets} (a,c) we show recent computations for the modification of bound states in the presence of thermal backgrounds. The results in (a) show the thermal evolution of the width and dispersion of the peaks in the spectral functions for the lowest lying mesons in two-dimensional QED at finite T~\cite{Barata:2025jhd}, demonstrating the rise of the imaginary part of the inter-fermionic potential~\cite{Brambilla:2016wgg,Brambilla:2010vq}. Figure (c) shows the evolution of thermalization time with drag in a open quantum system approach for meson evolution in a thermal state~\cite{Angelides:2025hjt}, see also~\cite{Blaizot:2017ypk}.

Turning the attention to jets, a lot of progress has been achieved using the Hamiltonian light-front formulation~\cite{Brodsky:1997de, Barata:2021yri,Barata:2022wim,Barata:2023clv,Qian:2024gph,Kreshchuk:2020kcz,Castro:2025ocx,Avramescu:2025lhr,Li:2025wzq,Li:2021zaw,Li:2020uhl,Li:2023jeh,Li:2024ufq} which is typically employed in perturbative approaches to jet quenching~\cite{Arnold:2019pxd,Baier:1996kr}. In panel (b) we show a result~\cite{Barata:2023clv} for the average transverse momentum squared acquired inside a jet when evolving in the presence of quark gluon plasma background. Energy loss, which dominates the evolution of jets and heavy flavor, can also be computed; this is illustrated in (d), where the total energy lost by a propagating heavy state is shown as a function of its speed, when immersed in a bulk matter in two-dimensional QED. This construction can be extended to the case where both the heavy probe and the bulk are fully dynamical, see~\cite{Barata:2025hgx}.

\begin{figure}[h!]
    \centering
    \includegraphics[width=0.6\linewidth]{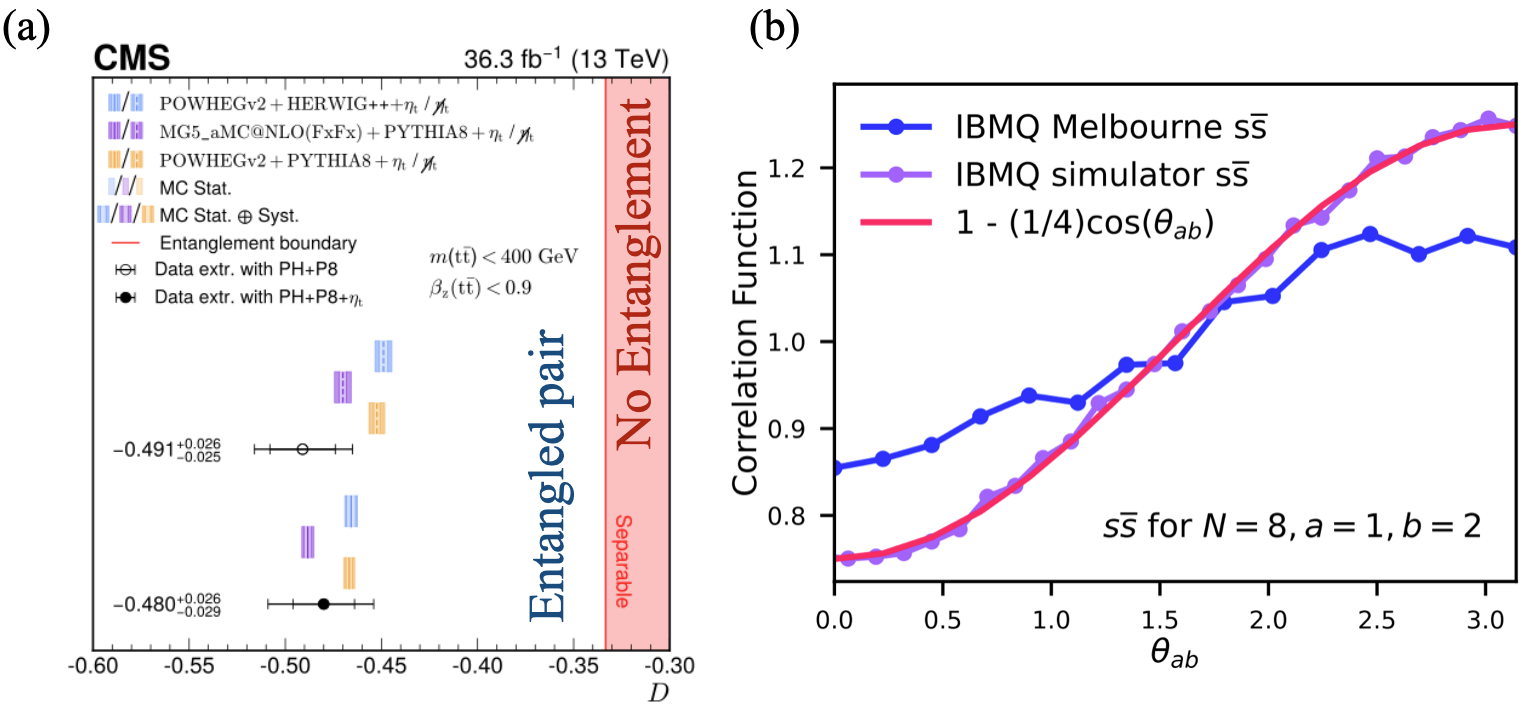}
    \caption{Tests of Bell inequalities in high energy physics: (a) measurement of $t\bar t$ correlations from the CMS collaboration~\cite{CMS:2024pts}, (b) spin correlations in a model calculation of $\Lambda \bar \Lambda$ hadronization in a QCD string~\cite{Gong:2021bcp}; curve shows expected analytic result, blue data was extracted from an IBM quantum processor, purple data used a classic emulator of a quantum computer.}
    \label{fig:spins}
\end{figure}

\noindent \paragraph{Spin correlations and entanglement tests:}
Finally, we want to highlight that QIS, beyond simulating quantum systems, can also assist in designing new measurements and observables which probe unique quantum features, see~\cite{Afik:2025ejh} for further discussion. Indeed in the context of high energy physics, entanglement measures have already been tested in $t\bar t$ production at the LHC~\cite{CMS:2024pts,ATLAS:2023fsd}. The experimental results from the CMS collaboration are shown in Fig.~\ref{fig:spins} (a), where we have made explicit that the data indicates the presence of genuine quantum correlations in measured $t\bar t$ pairs. 

Although such heavy particles are hard to produce in heavy collisions, entanglement tests can be performed in other particles where genuine quantum features are accessible. An interesting option are hyperons which, due to their self-analyzing decays, allow to reconstruct the associated spin structure. Work along these lines is being actively explored in the context of the future EIC, where a cleaner environment might allow to better reconstruct $\Lambda \bar \Lambda$ correlations, see~\cite{STAR:2025njp,Vanek:2024jai}. These might allow to clarify some of the open questions regarding polarization of these states; conversely, QIS methods also enable the test of current theoretical pictures by constructing models where spin correlations of hyperons can be simulated in real-time. This is illustrated Fig.~\ref{fig:spins} (b), where a model calculation of the spin correlation is shown as a function of the angle between the measured pair, using a real digital quantum computer~\cite{Gong:2021bcp,Barata:2023jgd}, see also e.g.~\cite{Afik:2025grr,vonKuk:2025kbv} for related studies. How such studies can be extended to the heavy ion context remains a open question, which might lead to interesting insights into the properties of QCD matter.

\section{Future Opportunities and Conclusion}\label{sec:conclusion}
To conclude, in this short summary we tried to highlight the possible connections between heavy ion physics and QIS. In particular, we have argued that many of the open questions about the bulk QCD matter being produced can, in principle, be studied using novel quantum technologies, which allow to explore finite density and real-time dynamics of quantum systems. We have exemplified some current applications in gauge theories which are of the interest of the heavy ion community, while recognizing the current limitations of quantum platforms. Several more aspects relevant for heavy ion physics which can and have been explored in QIS were not presented here; they cover topological structure~\cite{Funcke:2019zna,Ikeda:2023zil,Zache:2018cqq,Dempsey:2023gib}, equilibration and collective dynamics~\cite{Qian:2024xnr,Chen:2024pee,Janik:2025bbz,Zhou:2021kdl,Desaules:2022kse,Shao:2025ygy}, energy correlators~\cite{Barata:2025jhd, Barata:2024apg, Moult:2025nhu}, hadronization and string dynamics~\cite{Liu:2024lut,Batini:2024zst,Mallick:2024slg,Pichler:2015yqa,Surace:2024bht,Cochran:2024rwe}, particle scattering~\cite{Papaefstathiou:2024zsu,Davoudi:2025rdv,Belyansky:2023rgh}, among others.

\bibliography{Lib.bib}

\end{document}